\documentclass[final]{aipproc}
\layoutstyle{6x9}
\usepackage{bm}

\begin{document}

\title{Pomeron dynamics in the AdS space and structure functions of hadrons at small $\bm{x}$}



\author{Akira Watanabe}{
  address={Department of Physics, Tokyo University of Science, Shinjuku, Tokyo 162-8601, Japan}
}

\author{Katsuhiko Suzuki}{
  address={Department of Physics, Tokyo University of Science, Shinjuku, Tokyo 162-8601, Japan}
}

\begin{abstract}
The Pomeron dynamics is investigated via deep inelastic scattering (DIS) 
at small $x$
in the framework of holographic quantum chromodynamics.   
The small $x$ DIS process is assumed to be described by the graviton exchange between 
external vector current and hadron in the AdS space.  
Our calculations for  $F_2^p$, $F_2^\pi$, as well as the longitudinal counterpart $F_L^p$ are consistent with the experimental data.
We discuss origins of a difference between 
$F_2$ and $F_L$ in our approach.
\end{abstract}

\maketitle

\section{INTRODUCTION}

Deep inelastic lepton-nucleon scattering (DIS) at the small Bjorken-$x$ is a important tool to understand the origin of the Pomeron in quantum chromodynamics (QCD), which may be the multi-gluon exchange with the vacuum quantum number and provides dominant contributions  to various high energy scattering processes. 
Assuming the Pomeron exchange with the intercept $\alpha (0)$, one can express the structure function $F_2$ as $F_2 (x,Q^2) \sim x^{1- \alpha (0)}$ with the photon virtuality $Q^2$ and the Bjorken-$x$, 
$x = Q^2/s$.   
This expression is valid only in the small $x \ll 1$ region, where the Regge kinematics is satisfied. 
The experimental data~\cite{Breitweg:1998dz} indicate that the intercept 
depends on the resolution scale $Q^2$, namely, $\alpha (0) \sim 1.1$ at lower 
``soft'' $Q^2  < 1~\mbox{GeV}^2$ region, while it increases up to about 1.4 for higher ``hard'' $Q^2 
> 1~\mbox{GeV}^2$ domain.  
Hence, understanding of the Pomeron dynamics may reveal a ``missing'' link 
between non-perturbative and perturbative features of QCD.  
However, it is extremely difficult to calculate such a scale dependence theoretically, because the Pomeron intercept at the soft region, $Q^2 \sim 0$, is a highly  non-perturbative quantity.

In this brief report, based on our recent work \cite{Watanabe:2012uc}, we show calculations of the nucleon and pion structure functions within holographic QCD.  
By virtue of the gauge/string correspondence, the contribution to $F_2$ from the Pomeron exchange process could be identified with the Reggeized graviton exchange in the bulk AdS space. 
Corresponding scattering kernel was derived by Brower-Polchinski-Strassler-Tan (BPST) in Refs.~\cite{Polchinski:2002jw,Brower:2006ea,Cornalba:2007zb,Brower:2007qh,Brower:2010wf}.
In addition, the coupling of the target hadron ($h$) to the graviton ($g$) can be obtained by
calculating the $ghh$ three point function from the effective classical action~\cite{Abidin:2008hn}.  
Using these inputs we can systematically obtain the holographic description of the 
structure function at small $x$, instead of crude ``delta function approximation'' for the 
hadron density distributions adopted in Ref.~\cite{Brower:2010wf}.  
We discuss how the Pomeron intercept changes as the scale $Q^2$ increases from 
non-perturbative to perturbative regime.  

We also concentrate on the longitudinal structure function $F_L$, which exactly vanishes 
in the quark-parton model.  
Hence, in principle, $F_L$ is sensitive to the intrinsic dynamics and the interaction inside 
the hadron.  
We discuss possible origin of the longitudinal structure function at the small $x$ in the 
holographic QCD.

\section{BPST KERNEL AND POMERON-HADRON COUPLINGS}

The BPST Pomeron kernel $\chi$ was derived with a combination of higher dimensional string theory and the AdS/CFT correspondence in the region where the beta function was almost vanished. With this kernel, the structure function $F_2$ at the small $x$ is written as~\cite{Watanabe:2012uc,Brower:2010wf}
\begin{equation}
F_2(x,Q^2) = \frac{g_0^2 \rho^{3/2} Q^2}{32 \pi ^{5/2}} \int dzdz' P_{13}(z,Q^2) P_{24}(z') (zz') \ \mathrm{Im} [\chi(s,z,z') ] \ ,
\label{eq:F2}
\end{equation}
where $P_{13}$ and $P_{24}$ are ``overlap functions'' of the virtual photon and the target hadron with their fifth (or bulk) coordinates $z$ and $z'$.  
The BPST kernel contains two adjustable parameters to be determined later.

\begin{figure}[!b]
  \includegraphics[width=.99\textwidth]{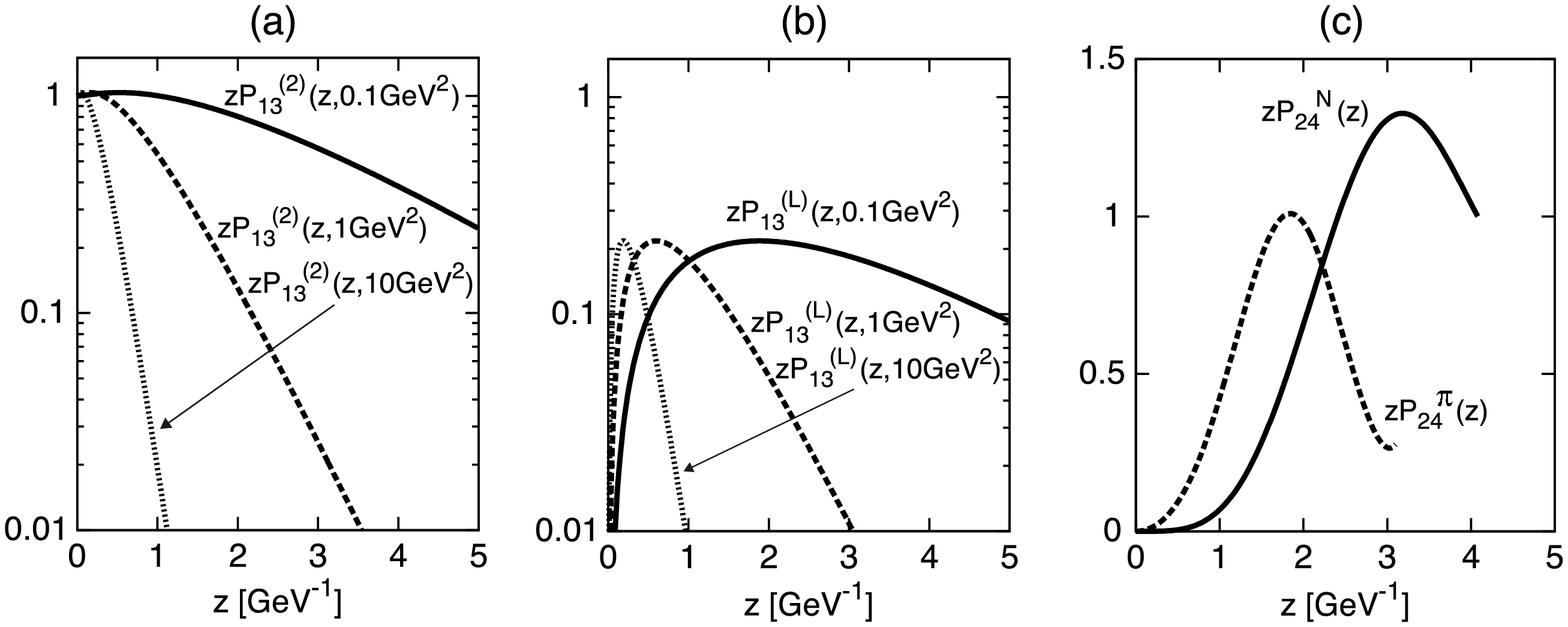}
  \caption{The overlap functions (a) $zP_{13}^{(2)}(z,Q^2)$ with $Q^2 = 0.1, 1, 10~\mbox{GeV}^2$ shown by solid, dashed, and dotted curves, respectively.
(b) $zP_{13}^{(L)}(z,Q^2) \equiv zP_{13}^{(2)}(z,Q^2) - zP_{13}^{(1)}(z,Q^2)$.  Notations are the same as those of (a),  
(c) $zP_{24}(z)$ for $N$ and $\pi$ shown by solid and dashed curves, respectively.  
}
  \label{fig:overlap_functions}
\end{figure}

For the overlap function of the virtual photon, we consider the wave function of the $R$-boson propagating in the AdS space and coupling to leptons on the boundary~\cite{Polchinski:2002jw,Brower:2010wf}.
We use $P_{13}^{(2)}(z,Q^2) = Q^2 z \left( K_0^2(Qz) + K_1^2(Qz) \right)$ to calculate $F_2$, and 
$P_{13}^{(1)}(z,Q^2) = Q^2 z K_1^2(Qz)$ for  $2xF_1$, which yield 
the longitudinal structure function $F_L = F_2 - 2xF_1$.

On the other hand, the hadronic overlap functions are shown to be density distribution functions in the bulk space, 
which can be extracted from the hadron-graviton-hadron three point functions~\cite{Abidin:2008hn}.
We rely on so called ``bottom-up'' approaches in holographic QCD, and assume 
that the nucleon is considered  as a normalizable mode of the 5D Dirac equation, and the pion emerges  
as a fluctuation of the bulk scalar field.
To obtain the three point functions, we introduce the metric perturbation, $\eta _{\mu \nu} \rightarrow \eta _{\mu \nu} + h_{\mu \nu}$, in the 5D classical action, and pick up the $h \psi \psi$ terms, where $\psi$ is the AdS wave function of the hadron~\cite{Watanabe:2012uc,Abidin:2008hn}.   
Here we use the ``hard-wall'' model where the AdS geometry is sharply cut off at the IR boundary, to break the conformal invariance to mimic the realistic QCD. 
We show calculated overlap functions in Fig.~\ref{fig:overlap_functions}, (a) $zP_{13}^{(2)}$,  
(b) $zP_{13}^{(L)}  \equiv zP_{13}^{(2)} - zP_{13}^{(1)}$, and (c) $zP_{24}$ for $N$, $\pi$.

\section{NUMERICAL RESULTS AND DISCUSSIONS}

For the numerical calculations, the parameters of the BPST kernel are fixed to reproduce the 
proton $F_2^p (x,Q^2)$ data~\cite{Aaron:2009aa}.  The hard-wall cutoff parameters for the nucleon 
and the pion are chosen to reproduce the nucleon mass and the pion decay constant.

\begin{figure}[!b]
  \includegraphics[width=.97\textwidth]{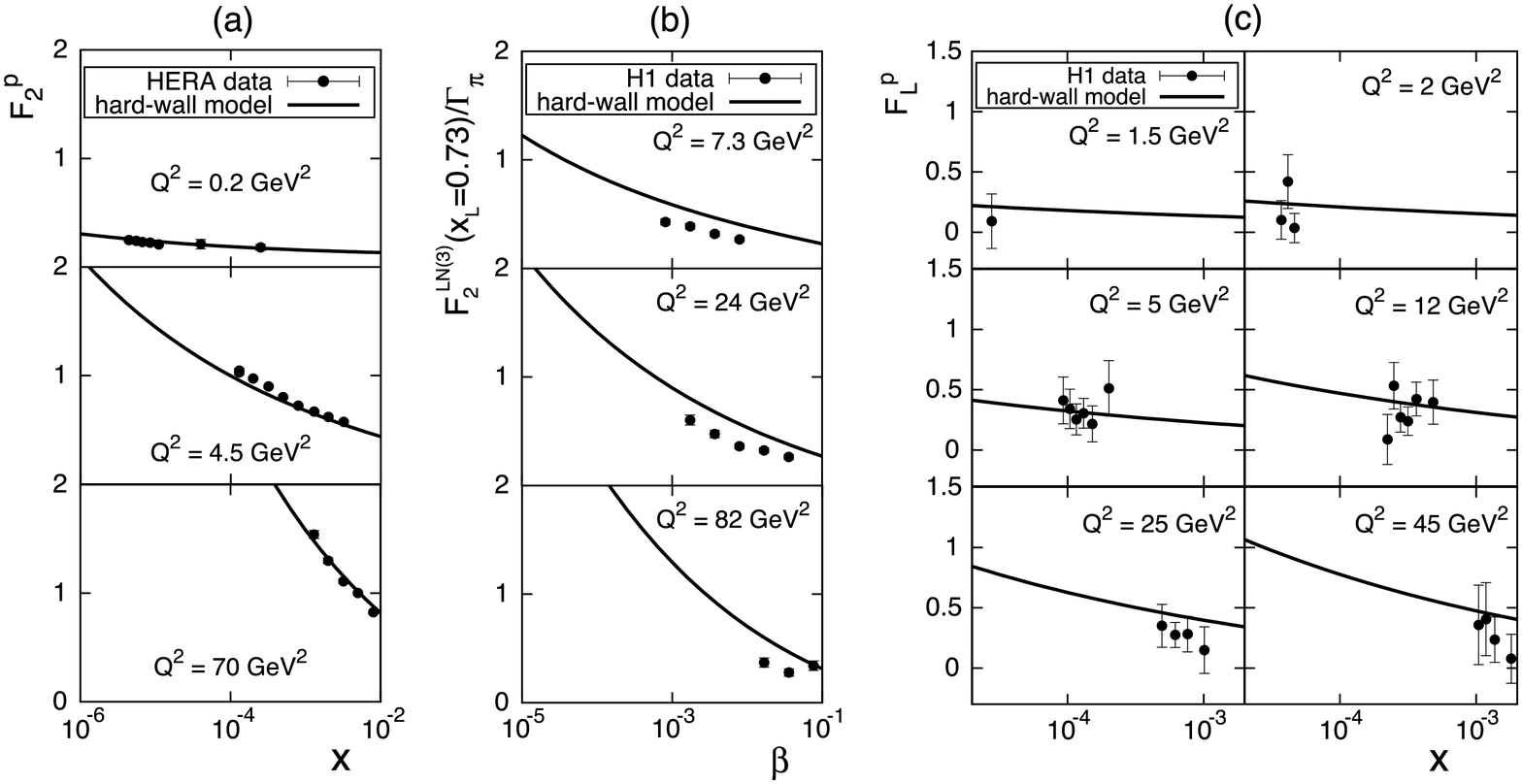}
  \caption{The nucleon and pion structure functions. The solid curves represent our calculations. 
\hspace{4mm}
(a) $F_2^p(x,Q^2)$ with the combined data by H1 and ZEUS~\cite{Aaron:2009aa}, 
(b) $F_2^{\pi}(\beta , Q^2)$ with the experimental data by H1 within the pion exchange model~\cite{Aaron:2010ab}, (c) $F_L^p (x,Q^2)$ with the H1 data~\cite{Collaboration:2010ry}}
    \label{fig:SFs}
\end{figure}

We show in Fig.~\ref{fig:SFs} the 
nucleon and pion structure functions, (a) $F_2^p$, (b) $F_2^\pi$, and 
(c) $F_L^p$.
Resulting  $F_2$ structure functions for the nucleon and the pion fairly agree with the experimental data,  although the calculation for the pion is based on the pion exchange 
model with certain theoretical ambiguities.
Our results for $F_L^p$ in the panel (c) are also in good agreement with the data.

\begin{figure}[!t]
  \includegraphics[width=.95\textwidth]{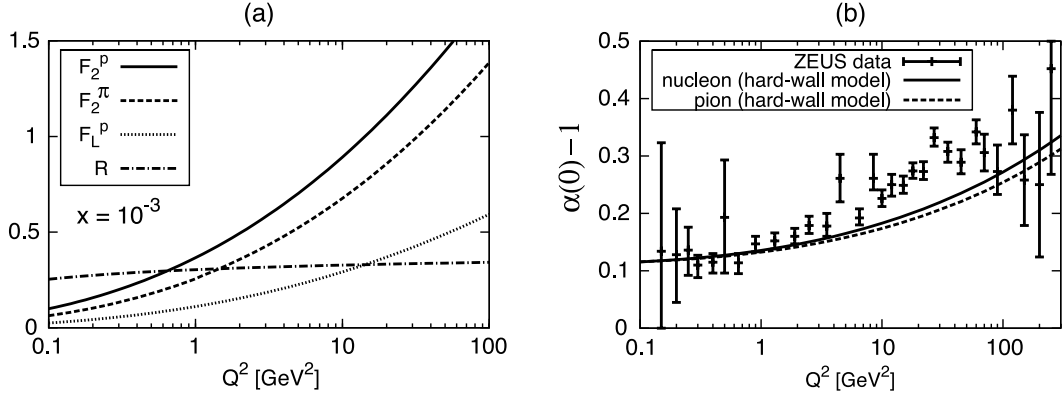}
  \caption{(a) $Q^2$ dependencies of $F_2^p$, $F_2^\pi$, $F_L^p$, and $R(x,Q^2)=F_L^p(x,Q^2)/F_2^p(x,Q^2)$ at $x=10^{-3}$.
\hspace{5mm}
(b) $Q^2$ dependence of the Pomeron intercept $\alpha (0)-1$.  Our results for the nucleon and 
the pion  are depicted by 
solid and dashed curves, respectively.  
The experimental data is taken from Ref.~\cite{Breitweg:1998dz}.}
  \label{fig:intercepts}
\end{figure}

In order to clarify the difference of $Q^2$ dependencies, 
we show in Fig.~\ref{fig:intercepts}~(a)  $F_2^p$, $F_2^\pi$, $F_L^p$, and a ratio $R=F_L^p/F_2^p$ at $x=10^{-3}$. 
The  $Q^2$ dependencies of the $F_2^p$ and  $F_2^\pi$ are similar as expected, although $F_2^p$ is somewhat larger than $F_2^\pi$ in absolute magnitude.    
Resulting ratio $R(Q^2)$ is found to be almost constant, and continues to rise very slowly as $Q^2$ increases.  
To understand this behavior qualitatively, we compare the photon overlap functions 
$P_{13}^{(2)}$ and $P_{13}^{(L)}$  
in Fig.~\ref{fig:overlap_functions}~(a) and (b).  
In Ref.~\cite{Watanabe:2012uc}, 
we have found the $Q^2$ dependence is closely related with the $z$-distributions of overlap 
functions.  
It seems clear that both 
$P_{13}^{(2)}$ and $P_{13}^{(L)}$  
show quite similar distributions in the bulk space, but  
$zP_{13}^{(L)}$ is significantly suppressed in magnitude.  
These facts cause the longitudinal to transverse  ratio $R$ to be almost 
independent of $Q^2$ and much less than 1. 
However, our result 
is incompatible with the prediction of perturbative QCD, $R(x,Q^2) \sim 1/\log Q^2$ 
for large $Q^2$. 
This may suggest that our framework is inadequate to the description of the high $Q^2$ region.

We finally show in Fig.~\ref{fig:intercepts}~(b) the $Q^2$ dependence of the Pomeron intercept.  
The intercept $\alpha (0)$ is about 1.1 for the non-perturbative $Q^2$ region, and reaches to 
about 1.3 at perturbative larger $Q^2$ scale.  
The essential characters of both soft and hard Pomerons are well reproduced in 
our model.  
We also note that 
the similarity of the proton and pion results 
indicates the universality of the Pomeron intercept, which is in fact realized in the 
observations.

In conclusion, we have calculated the nucleon and pion structure functions at the small $x$ in the framework of holographic QCD, which are in good agreement with the data.
Our framework can be applied to any kind of hadrons without any adjustable parameter, by calculating the overlap function of the hadron in the bulk space.  
We have reproduced the non-trivial $Q^2$ dependence of the Pomeron intercept, which 
is nothing but an unified description of the soft and hard Pomerons in our model.

\bibliographystyle{aipproc}   

\begin{thebibliography}{99}


\bibitem{Breitweg:1998dz}
  J.~Breitweg {\it et al.}  (ZEUS Collaboration),
  Eur.\ Phys.\ J.\  C {\bf 7}, 609 (1999).


\bibitem{Watanabe:2012uc} 
  A.~Watanabe and K.~Suzuki,
  Phys.\ Rev.\ D {\bf 86}, 035011 (2012).


\bibitem{Polchinski:2002jw} 
  J.~Polchinski and M.~J.~Strassler,
  JHEP {\bf 0305}, 012 (2003).


\bibitem{Brower:2006ea} 
  R.~C.~Brower, J.~Polchinski, M.~J.~Strassler and C.~-I.~Tan,
  JHEP {\bf 0712}, 005 (2007).


\bibitem{Cornalba:2007zb} 
  L.~Cornalba, M.~S.~Costa and J.~Penedones,
  JHEP {\bf 0709}, 037 (2007).


\bibitem{Brower:2007qh} 
  R.~C.~Brower, M.~J.~Strassler and C.~-I.~Tan,
  JHEP {\bf 0903}, 050 (2009);~{\bf 0903}, 092 (2009).
%


\bibitem{Brower:2010wf} 
  R.~C.~Brower, M.~Djuric, I.~Sarcevic and C.~-I.~Tan,
  JHEP {\bf 1011}, 051 (2010).


\bibitem{Abidin:2008hn} 
  Z.~Abidin and C.~E.~Carlson,
  Phys.\ Rev.\ D {\bf 77}, 115021 (2008);~{\bf 79}, 115003 (2009).
%


\bibitem{Aaron:2009aa}
  F.~D.~Aaron {\it et al.}  (H1 and ZEUS Collaborations),
  JHEP {\bf 1001}, 109 (2010).


\bibitem{Aaron:2010ab} 
  F.~D.~Aaron {\it et al.}  (H1 Collaboration),
  Eur.\ Phys.\ J.\ C {\bf 68}, 381 (2010).


\bibitem{Collaboration:2010ry} 
  F.~D.~Aaron {\it et al.}  (H1 Collaboration),
  Eur.\ Phys.\ J.\ C {\bf 71}, 1579 (2011).


\end{thebibliography}

\end{document}